\documentclass[12pt]{article}
\usepackage{epsfig}
\usepackage{subfigure}
\usepackage{amsmath}
\usepackage{amsfonts}
\usepackage{amssymb}
\usepackage{url}
\usepackage{hyperref}
\newcommand{\be}{\begin{equation}}
\newcommand{\ee}{\end{equation}}
\newcommand{\bea}{\begin{eqnarray}}
\newcommand{\eea}{\end{eqnarray}}

\begin{document}
%%%%%%%%%%%%%%%%%%%%%%%%%%%%
\begin{titlepage}
%%%%% PREPRINT NUMBERS %%%%%%
\begin{flushright}
\today
\end{flushright}
\vspace{4\baselineskip}
%%%%%%%%%%%%%%%%%%% TITLE %%%%%%%%%%%%%%%%%%
\begin{center}
{\Large\bf 
A solution to the little hierarchy problem in\\
a partly ${\cal N}=2$ extension of the MSSM
}
\end{center}
%%%%%%%%%%%%%%%% AUTHORS %%%%%%%%%%%%%%%%%%%%%%%
\vspace{1cm}
\begin{center}
{\large
Tatsuru Kikuchi
\footnote{\tt E-mail:tatsuru@post.kek.jp}
}
\end{center}
%%%%%%%%%%%%%%%%%%%%%%% AFFILIATION %%%%%%%%%%%%
\vspace{0.2cm}
\begin{center}
{\small \it Theory Division, KEK,
Oho 1-1, Tsukuba, Ibaraki, 305-0801, Japan}\\
\medskip
\vskip 5mm
\end{center}
\vskip 5mm
\begin{abstract}
We extend a model of the Dirac gauginos, which originate from 
${\cal N} =2$ supersymmetry (SUSY) for the gauge sector,
such that the ${\cal N} =2$ SUSY is imposed also to the sfermion sector
but only for the 3rd generation squarks and sleptons.
In addition to the ${\cal N} =2$ supersymmetry, our model is constructed
based on the $SU(3)_c \times SU(3)_L \times U(1)^\prime$ gauge symmetry.
By this extension, the dominant source of radiative correction
to the Higgs mass squared coming from the stop loop becomes controllable
based on the enhanced symmetry.
Then it becomes a viable model which provides a solution to 
the little hierarchy problem in SUSY models.
Even in the original scenario, the Dirac gauginos can be superheavy,
of order $10$ TeV or so, while keeping the scalar masses at the weak scale.
This possibility is phenomenologically interesting because it can suppress 
the unwanted flavor changing processes.
And in scope of the LHC, this scenario can have a very distinct signature 
related to the exotic sfermions which are accompanied as the ${\cal N} =2$
superpartners as well as the Dirac gauginos.
\end{abstract}
\end{titlepage}
%%%%%%%%%%%%%%%%%%%%%%%%%%%%%%%%%%%%%%%%%%%%%%%%
\section{Introduction}
%%%%%%%%%%%%%%%%%%%%%%%%%%%%%%%%%%%%%%%%%%%%%%%%
Supersymmetry (SUSY) extension is one of the most promising way 
 to solve the gauge hierarchy problem in the standard model \cite{SUSY}. 
However, since any superpartners have not been observed 
 in current experiments, SUSY should be broken at low energies. 
Furthermore, soft SUSY breaking terms are severely constrained 
 to be almost flavor blind and CP invariant. 
Thus, the SUSY breaking has to be mediated to the visible sector 
 in some clever way not to induce too large CP and flavor violation effects. 
Some mechanisms to achieve such SUSY breaking mediations  
 have been proposed \cite{Luty:2005sn}. 

On the other hand, in view of the LHC collider physics,
it is quite important to distinguish if the gauginos are Dirac or Majorana.
So far many studies have carried out based on the assumption
that the gauginos are Majorana particles as in the MSSM.
However, there exist some models which give rise to the gaugino
as Dirac particle which can naturally realized in models
with $U(1)_R$ symmetry \cite{Randall}.
For example, supersoft SUSY breaking scenario \cite{Fox:2002bu, Chacko:2004mi}
which has partly ${\cal N}=2$ supersymmetry in gauge sector
provide a model with Dirac gauginos.
Some other extensions of the model and the related phenomenology
can be found in the literature \cite{Nelson:2002ca, Nomura:2005rj}.
Supersoft SUSY breaking scenario also has some good features in
regards to the SUSY CP problems or the EDMs \cite{Hisano:2006mv}.

In this letter, we extend a model of the Dirac gauginos, which originate from 
${\cal N} =2$ SUSY for the gauge sector,
such that the ${\cal N} =2$ SUSY is imposed also to the sfermion sector
but only for the 3rd generation squarks and sleptons.
By this extension, the dominant source of radiative correction
to the Higgs mass squared coming from the stop loop becomes controllable
based on the enhanced symmetry.
Then it becomes a viable model which provides a solution to 
the little hierarchy problem in SUSY models.
Even in the original scenario, the Dirac gauginos can be superheavy,
of order $10$ TeV or so, while keeping the scalar masses at the weak scale.
This possibility is phenomenologically interesting because it can suppress 
the unwanted flavor changing processes.
And in scope of the LHC, this scenario can have a very distinct signature 
related to the exotic sfermions which are accompanied as the ${\cal N} =2$
superpartners as well as the Dirac gauginos.

%%%%%%%%%%%%%%%%%%%%%%%%%%%%%%%%%%%%%%%%%%%%%%%%
\section{Dirac gaugino in a warped extra dimension}
%%%%%%%%%%%%%%%%%%%%%%%%%%%%%%%%%%%%%%%%%%%%%%%%
\subsection{Introduction to the model of Dirac gaugino}
Here we give a brief description of the Dirac gaugino model based on the ${\cal N} = 2$ SUSY in gauge sector.
At first, an ${\cal N} = 2$ vector multiplet ${\cal V} = (V,~\Sigma)$ contains an ${\cal N} = 1$ vector multiplet,
$V= (A^\mu,~\lambda)$ and an $N = 1$ chiral multiplet in the adjoint representation, $\Sigma = (\phi,~\psi)$.
Basically, in the Dirac gaugino model, the gaugino ($\lambda$) has the Dirac mass terms together with
an adjoint fermion ($\psi$). Suppose the adjoint fermion has
mass term $m_\Sigma$, the mass matrix for these states are given by
\be
{\cal L}
=
\left(
\begin{array}{cc}
\lambda & \psi
\end{array}
\right)
\left(
\begin{array}{cc}
0 & M_{D} \\
M_{D} & m_{\Sigma} \\
\end{array}
\right)
\left(
\begin{array}{c}
\lambda \\ \psi
\end{array}
\right) \;.
\ee
Therefore, in the mass eigenstates, there are two copious
gauginos for each gauge group: 
$(\tilde{g}_1,~\tilde{g}_2)$ for SU(3),
$(\tilde{W}_1,~\tilde{W}_2)$ for SU(2), and
$(\tilde{B}_1,~\tilde{B}_2)$ for U(1).
Many studies on the nature of the Dirac gauginos
in view of the collider physics have been performed.
Some studies on the Dirac nature of either charginos or neutralinos, 
which are some mixed states of both $(\tilde{W}_1,~\tilde{W}_2)$ and $(\tilde{B}_1,~\tilde{B}_2)$,
are given in \cite{Choi:2001ww, Aguilar-Saavedra:2003hw, Choi:2003fs, Hagiwara:2005ym} 
and a study on the exotic gluinos $(\tilde{g}_1,~\tilde{g}_2)$ having Dirac mass terms
in \cite{Nojiri:2007jm}.

\subsection{Dirac gaugino masses in a warped extra dimension}
We consider a SUSY model 
 in the warped five dimensional brane world scenario. 
The fifth dimension is compactified on the orbifold $S^1/Z_2$ 
 with two branes, ultraviolet (UV) and infrared (IR) branes, 
 sitting on each orbifold fixed point. 
With an appropriate tuning for cosmological constants 
 in the bulk and on the branes, 
 we obtain the warped metric, 
\begin{eqnarray}
 d s^2 = e^{-2 k r_c |y|} \eta_{\mu \nu} d x^{\mu} d x^{\nu} 
 - r_c^2 d y^2 \; , 
\end{eqnarray}
 for $-\pi\leq y\leq\pi$, where $k$ is the AdS curvature, and 
 $r_c$ and $y$ are the radius and the angle of $S^1$, respectively.
The most important feature of the warped extra dimension model 
 is that the mass scale of the IR brane is warped down to 
 a low scale by the warp factor, $\omega = e^{-k r_c \pi}$, 
 in four dimensional effective theory. 
For simplicity, we take the cutoff of the original five dimensional theory 
 and the AdS curvature as 
 $M_5 \simeq k \simeq M_P=2.4 \times 10^{18}$ GeV, 
 the four dimensional Planck mass,
 and so we obtain the effective cutoff scale 
 as $\Lambda_{\rm IR}= \omega M_P$ in effective four dimensional theory. 
Now let us take the warp factor so as for the GUT scale 
 to be the effective cutoff scale 
 $ M_{\rm GUT}= \Lambda_{\rm IR}=\omega M_P$, 
 namely $\omega \simeq 0.01$.

When we start from the warped five dimensional setup,
and introduce an extra U(1) gauge multiplet
which is localized on the IR brane,
the ${\cal N}=2$ vector multiplet consists of
$(A_M,~\lambda_{1,2},~\Sigma)$ where
$\Sigma$ is an adjoint scalar multiplet.

In principle, it is always allowed
to have the following operator that generate the Dirac
gaugino masses:
\be
{\cal L}
= \int d^2 \theta \frac{\Sigma}{M_5 \omega} 
(W_{U(1)'})^\alpha (W_{\rm MSSM})_\alpha 
+ h.c. \;,
\ee
where the warp factor is assigned
for the rescaling of $\Sigma$, $\Sigma \to \Sigma/\omega$. 
After developing the VEV of the D-term in $W_{U(1)'}$,
the resultant Dirac gaugino masses at the IR scale 
or the GUT scale
after rescaling $\left< D' \right> \to \omega^2 \left< D' \right>$ are given by
\be
M_{D_i}= \alpha_{\rm GUT}^{1/2}  \frac{\omega \left< D' \right>}{M_5} \;.
\ee
Below the IR scale the Dirac gaugino masses at a given scale 
$\mu$ are given by
\be
M_{D_i} (\mu) = \left(\frac{\alpha_i(\mu)}{\alpha_{\rm GUT}} 
\right)^{\frac{b_i - 2 c_i}{2 b_i}} M_{D_i} \;,
\label{RG}
\ee
where $b_i$ is the beta function and
$c_i$ is the quadratic Casimir of the adjoint representation.

Given a gaugino mass $M_\lambda$, scalar masses
are arisen at the one loop level:
\be
m^2_{\rm supersoft} = \frac{c_i \alpha_i}{\pi} 
\ln \left(4 %+ \frac{m_{\Sigma}^2}{M_{D_i}^2} 
\right) M_{D_i}^2 \;,
\ee
where we took a limit of $m_{\Sigma} \ll M_{D_i}$. 
Hence the scalar masses are suppressed by
a factor of $\frac{\alpha_i}{\pi}$ compared to the gaugino masses.
For instance, in order to have scalar mass scale of order 
$m_{\rm scalar} \sim 100$ GeV,
the corresponding gaugino masses need to be around $M_{D_i} \sim 10$ TeV.
Remarkable feature of the supersoft SUSY breaking scenario is that
the scalar masses generated by gaugino loops are finite, and it never
receive renormalization except for the gaugino mass itself,
though the gaugino masses are renormalized by the amount of gauge 
couplings as in Eq. (\ref{RG}).

In additional to the above gaugino mass terms, there exist another 
contribution to the gaugino masses from anomaly mediation,
which are the Majorana terms and not the Dirac mass terms.
\be
M_i^{\rm AMSB} = \frac{b_i g_i^2}{16 \pi^2} F_\phi \;,
\ee
where $F_{\phi}$ is the F-term of the conpensator multiplet, 
$\phi = 1 + \theta^2 F_{\phi} \simeq 1 + \theta^2 m_{3/2}$.
Since the supersoft SUSY breaking contribution gives masses
of order $10$ TeV in order to have a scalar masses,
this contributions can be negligible in most cases.

Note that the gravitino mass is not an independent quantity it is fixed
so as to cancel the cosmological constant. 
To obtain the vanishing cosmological constant, we need to have an appropriate 
constant superpotential on the UV brane.
The condition for vanishing cosmological constant is described as
\be
\left<V \right> = \omega^4 \left< D' \right>^2 - 3 \frac{|W|^2}{M_p^2} = 0 \;,
\ee
where we took into account of the rescaling 
$\left< D' \right> \to \omega^2 \left< D' \right>$,
the resultant superpotential is 
$|W|^2 = \left(\omega^4 \left< D' \right>^2 M_p^2 \right)/3$, 
and then the gravitino mass is given by
\be
m_{3/2} = \frac{|W|}{M_p^2} = \frac{\omega^2 \left< D' \right>}{\sqrt{3} M_p} 
\simeq \frac{\omega M_{D_i}}{\sqrt{3}}
\sim m_{\rm supersoft}\;,
\ee
where we took a warp factor as $\omega \simeq 0.01$.
Hence the anomaly mediated contribution becomes well negligible.

\subsection{Radiative Electroweak Symmetry Breaking}
In the MSSM, the upper bound on the lightest Higgs mass 
at one loop level is given by
\be
m_{h}^2 \lesssim
m_Z^2 \cos^2 2\beta + \frac{3 v^2 y_t^4}{4 \pi^2}
\sin^4 \beta
\ln \left(\frac{m_{\tilde{t}}^2}{m_t^2} \right) \;.
\ee
In order to satisfy the LEP-II experimental lower bound on
the Higgs mass $m_h \gtrsim 114$ GeV, we need to push up
the stop mass $m_{\tilde{t}} \gtrsim 1$ TeV which may cause
a destabilization of the gauge hierarchy, that is 
the so called little hierarchy problem.

On the other hand, in the supersoft SUSY breaking scenario,
the leading contribution to the negative Higgs boson mass squared
comes from top quark and squark loop, which is given by
\be
m_{H_u}^2 \simeq
m_{\tilde{\ell}}^2-  \frac{3 y_t^2}{4 \pi^2} m_{\tilde{t}}^2 
\ln \left(\frac{M_{D_3}}{m_{\tilde{t}}}\right)\;,
~~m_{H_d}^2 \simeq m_{\tilde{\ell}}^2 \;.
\ee
Hence, the mediation scale in this scenario is not high but just 
$M_{\rm mess} = M_{D_3} \cong 10$ TeV, the fine-tuning is relaxed 
compared to the minimal SUGRA or any other high scale SUSY breaking
scenario.
Putting the results all together, the negative Higgs boson mass squared
parameter is written by
\be
m_{H_u}^2 \simeq
m_{\tilde{\ell}}^2- 
\frac{3 y_t^2 \alpha_3^{35/18} \alpha_{\rm GUT}^{-4/9}\ln \left(4 \right)}
{4 \pi^3} \frac{\omega \left< D' \right>}{M_5}
\ln \left(\frac{M_{D_3}}{m_{\tilde{t}}}\right)\;,
~~m_{H_d}^2 \simeq m_{\tilde{\ell}}^2 \;.
\ee
For a given Higgs boson mass squared parameter,
the electroweak symmetry breaking condition is described by
\be
\frac{M_Z^2}{2} = - m_{H_u}^2 - |\mu|^2\;.
\ee
So, if we take $|\mu| \cong m_{\tilde{\ell}}$,
the correct electroweak symmetry breaking can be achieved
without requiring fine-tuning of the Higgs mass parameters.

Therefore, the important point to solve the little hierarchy problem
in this scenario is to raise the stop mass
while keeping the slepton mass at the weak scale.
Indeed, there exists a mass hierarchy between the stop 
and the slepton according to the gauge couplings or 
the gauge quantum numbers.
\be
\frac{m_{\tilde{t}}^2}{m_{\tilde{\ell}}^2}
\sim \frac{c_3 \alpha_3}{c_2 \alpha_2} 
\times \frac{M_{D_3}^2}{M_{D_2}^2} \;.
\ee
Hence, we can take $m_{\tilde{t}} \simeq 1$ TeV
and $m_{\tilde{\ell}} \simeq 100$ GeV at the same time,
which is needed to solve the little hierarchy problem.

%%%%%%%%%%%%%%%%%%%%%%%%%%%%%%%%%%%%%
\section{A partly ${\cal N}=2$ extension of the MSSM}
%%%%%%%%%%%%%%%%%%%%%%%%%%%%%%%%%%%%%
\subsection{Introduction to the model of partly ${\cal N}=2$ extended MSSM}
We extend a model of the Dirac gauginos, which originate from 
${\cal N} =2$ SUSY for the gauge sector,
such that the ${\cal N} =2$ SUSY is imposed also to the sfermion sector
but only for the 3rd generation squarks and sleptons.
\begin{table}[htbp]
\begin{center}
\begin{tabular}{|c|c|c|c|}
\hline
& 1st generation & 2nd generation & 3rd generation $\equiv$ $SU(2)_{\cal R}$ doublet \\
\hline \hline
quark doublet & $Q_1$ & $Q_2$ & ${\cal Q} = (q_3,~\overline{q}_3)$ \\
\hline
singlet quark & $u^c_1$ & $u^c_2$ &${\cal U}^c = (u^c_3,~\overline{u}^c_3)$ \\
\hline 
singlet quark & $d^c_1$ & $d^c_2$ &${\cal D}^c = (d^c_3,~\overline{d}^c_3)$ \\
\hline \hline
\end{tabular}
\end{center}
\caption{Chiral multiplets in the partly ${\cal N}=2$ extension of the MSSM}
\end{table}
\begin{table}[htbp]
\begin{center}
\begin{tabular}{|c|c|}
\hline
 gauge group &  vector multiplet \\
\hline \hline
SU(3) & ${\cal V}^A = (V^A,~\Sigma^A)~(A=1,\cdots, 8)$ \\
\hline
SU(2) & ${\cal V}^a = (V^a,~\Sigma^a)~(a=1,2,3)$ \\
\hline 
U(1) & ${\cal V} = (V,~\Sigma)$ \\
\hline \hline
\end{tabular}
\end{center}
\caption{Vector multiplets in the partly ${\cal N}=2$ extension of the MSSM}
\end{table}
Now we evaluate the effects of the Yukawa interactions between the MSSM matter
in the 3rd generation and the exotic states added in the 3rd generation
to the Higgs mass squared.

In order to have Higgs doublet from the adjoint field, which is almost the same situation
as occurred in the Littlest Higgs Model, we have to extend the SM gauge group, 
$SU(3)_c \times SU(2)_L \times U(1)_Y$ to $SU(3)_c \times SU(3)_L \times U(1)^\prime$,
and $SU(3)_L \times U(1)^\prime$ will be broken down to the Standard $SU(2)_L \times U(1)_Y$.
The adjoint field contains two Higgs doublet in addition to the triplet and singlet representations.
\be
\Sigma =
\left(
\begin{array}{cc}
 T & H_u \\
\overline{H}_u & S \\
\end{array}
\right) \;,
\ee
where $T$ is triplet, $H_u$ and $\overline{H}_u$ are doublet, and $S$ is singlet.
On the other hand, quark doublet is contained in the $SU(3)_L$ fundamental representation
such that
\be
Q_3 = \left(
\begin{array}{c}
q_3 \\
u^c_3
\end{array}
\right) \;,
\ee
and also we would have ${\cal N}=2$  partner of it:
\be
\overline{Q}_3 = \left(
\begin{array}{c}
\overline{q}_3 \\
\overline{u}^c_3
\end{array}
\right)\;.
\ee
The original interaction which provides a top Yukawa coupling is originated from
gauge interaction because of ${\cal N}=2$ supersymmetry.
\bea
{\cal L} &=& \int d^4 \theta \,g  \left[
\overline{Q}_3^\dag \,\Sigma \,\overline{Q}_3
+ Q_3^\dag \, \Sigma \,Q_3
\right]
\nonumber\\
&=&  \int d^2 \theta \, g \,\left[
\left(
\overline{q}_3 H_u \overline{u}^c_3
+ \overline{q}_3 \overline{H}_u \overline{u}^c_3  + \cdots \right) 
+ \left(
q_3 H_u u^c_3 + q_3 \overline{H}_u u^c_3  + \cdots \right) \right] \;.
\eea
The effective theory which we analyze below the GUT scale is the MSSM 
with the right-handed neutrinos. The low energy effective superpotential in this model is given by
\begin{eqnarray}
W_{\rm eff} &=&  \sum_{i,j = 1,2} \left(
Y_{u}^{ij} u^c_i q_j H_u 
+ Y_d^{ij} d^c_i q_j H_d
+ Y_{\nu}^{ij} \nu^c_i L_j H_u 
+ Y_e^{ij} e^c_i L_j H_d
+ \frac{1}{2} M_{R_{ij}} \nu^c_i  \nu^c_j \right)
\nonumber\\
&+&  g \,\left[
\left(
\overline{q}_3 H_u \overline{u}^c_3
+ \overline{q}_3 \overline{H}_u \overline{u}^c_3  + \cdots \right) 
+ \left(
q_3 H_u u^c_3 + q_3 \overline{H}_u u^c_3  + \cdots \right) \right]
\nonumber\\
&+& \mu H_d  H_u \;,
\label{Yukawa4}
\end{eqnarray} 
where the 2nd line describes the new interactions which appear
only in the partly ${\cal N}=2$ extension of the MSSM.

Interestingly, in the model of partly ${\cal N}=2$ extension of the MSSM, 
in contrast to the usual MSSM,  the lightest Higgs mass receives no divergent contribution
from the stops in the loop, and it becomes finite.
That is basically because ${\cal N} = 2$ partners of the top quarks,
$(\overline{q}_3,\,\overline{u}_3^c)$ can play the role to cancel the divergence
from the top quarks in the loop for the radiative corrections to the Higgs mass squared.
In this sense, the Higgs mass squared coming from the stop loop becomes controllable
based on the enhanced symmetry, and it can provides a natural solution to 
the little hierarchy problem in the MSSM.

In this setup, the upper bound on the lightest Higgs mass 
at one loop level is given by
\be
m_{h}^2 \lesssim
m_Z^2 \cos^2 2\beta + \frac{3 v^2 g^2}{4 \pi^2}
\sin^4 \beta
\left[
\ln \left(\frac{m_{\overline{t}_L}^2}{m_{t_L}^2} \right) 
+
\ln \left(\frac{m_{\overline{t}_R}^2}{m_{t_R}^2} \right) 
\right] \;.
\ee
So, in the limit of exact ${\cal N} = 2$ supersymmetry,
$m_{\overline{t}_L} = m_{t_L}$ and $m_{\overline{t}_R} = m_{t_R}$,
we end up with exactly the finite Higgs boson mass
as expected.

\newpage
\section*{Acknowledgments}
The work of T.K. is supported by the Research
Fellowship of the Japan Society for the Promotion of Science (\#1911329).

%%%%%%%%%%%%%%%%%%%%%%%%%%%%%%%%%%%%%%

\end{document}